# Visibility predicts priming within but not between people: a cautionary tale for studies of cognitive individual differences.

Frederic Boy & Petroc Sumner

School of Psychology, Cardiff University

Tower building / Park place

Cardiff, CF10 3 AT, United Kingdom

**Abstract**:

With resurgent interest in individual differences in perception, cognition and behavioural control, as early indicators of disease, endophenotypes, or a means to relate brain structure to function, behavioural tasks are increasingly being transferred from within-subject settings to between-group or correlational designs. The assumption is that where we know the mechanisms underlying within-subject effects, these effects can be used to measure individual differences in those same mechanisms. However, between-subject variability can arise from an entirely different source from that driving within-subject effects, and here we report a clear-cut demonstration of this. We examined the debated relationship between the visibility of a masked-prime stimulus and the direction of priming it causes (positive or reversed). Such reversal of priming has been hypothesized to reflect an automatic inhibitory mechanism that controls partially activated responses and allows behavioural flexibility. Within subjects, we found an unambiguous systematic transition from reversed priming to positive priming as prime visibility increased, replicated seven times, and using different stimulus manipulations. However, across individuals there was never a relationship between prime discrimination ability and priming. Specifically, these data resolve the controversial debate on visibility and reversed priming, indicating that they arise from independent processes relying on partially shared stimulus signals. More generally, they stand as an exemplar case in which variance between individuals arises from a different source from that produced by stimulus manipulations.

1. Introduction

Psychology has always contained within in it a division between two approaches (e.g. Hull, 1945; Cronbach, 1957). One seeks to assesses and explain differences between individuals, normally via correlational methods, while the other investigates basic cognitive processes with experiments that treat individual differences as nuisance variation. Cronbach (1957) hoped that 'the two disciplines of psychology' would converge and integrate because, he argued, 'kept independent, they can give only wrong answers or no answers at all regarding certain important problems' (p. 673).

It is likely that Cronbach would be disappointed with the degree of integration achieved more than 50 years later, but a new impetus for integration is now being driven from a direction that Cronbach might not have anticipated - psychological medicine, imaging and genetics, where there is increasing use of behavioural tasks from experimental psychology to measure individual differences in perception, cognition and behavioural control. To take three examples, there is the hope in psychiatric genetics of finding cognitive 'endophenotypes' – heritable and stable differences in cognitive mechanisms associated with psychiatric illness (Gottesman & Gould, 2003); there is growing endeavour in brain imaging to relate individual differences in structure, such as white matter connectivity, to differences in function; and there is accelerating interest in ageing, which is most easily studied cross-sectionally rather than longitudinally.

**The problem for integration.**

In these examples, it appears to be commonly assumed that where a task has been used successfully to reveal and investigate specific cognitive mechanisms through specific manipulations of stimuli or conditions, that task can then be simply used to measure how people differ in those mechanisms. In other words, that established within-subject phenomena will have easily-interpreted translation to individual differences. Unfortunately this is not true. Variance between individuals can arise from an entirely different source from that driving, and therefore studied using, within-subject effects (e.g. Borsboom, Kievit, Cervone & Hood, 2009). As we will show below, this can happen not just for complex individual differences such as IQ or personality, but even when a task is supposed to tap a much more basic mechanism, where it is intuitive to assume that individual differences would come from that same basic mechanism.

The theoretical basis for the difficulty in integrating correlational and experimental approaches is that, 'barring perhaps the most basic laboratory tasks for which assumptions like ergodicity or measurement invariance over individuals might be taken to hold true, any theory on intra-individual processes is compatible with any theory of inter-individual differences' (Borsboom, Kievit, Cervone & Hood, 2009, p 19). In other words, without additional simplifying assumptions, it is not possible to infer anything about within-person dependence from between-person comparisons, and vice versa. This problem is illustrated in Figure 1, which shows four hypothetical examples of how within-subject variance may or may not align with between-subject variance.

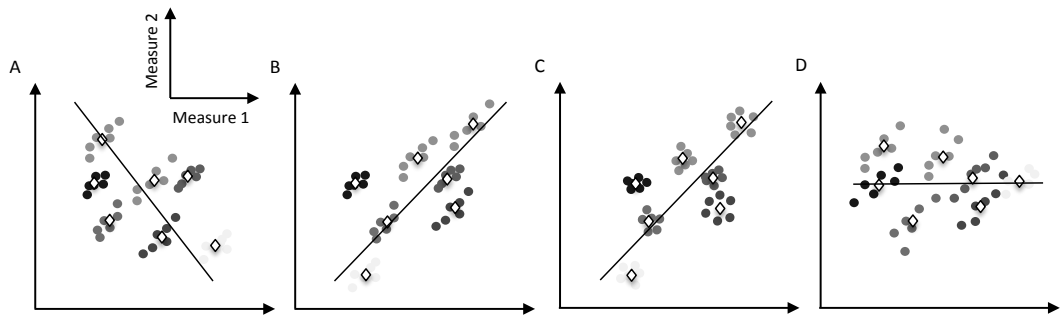

Figure 1: Schematic illustration of the theoretical independence of within- and between-subject variance through four examples. A) Within each individual (different gray tones), measurements 1 and 2 covary positively with each other, while individuals' means (white diamonds) covary negatively across the sample. This situation may appear counter-intuitive, but we are familiar with it in many circumstances with bidirectional causality: e.g. clothing thickness will correlate positively with feelings of warmth if clothing is manipulated within individuals; but across individuals, people who on average feel colder are likely to put thicker clothes on. B) Within each individual, measurements 1 and 2 are positively correlated, and individual's means also covary positively across the sample, as intuitively expected in many situations C) Within each individual, measurements 1 and 2 are not correlated, but individual's means covary positively across the sample. D) Within each individual, measurements 1 and 2 are correlated (positively in this example), but the means do not covary across the sample.

**Multiply-determined traits vs. basic mechanisms**

Counterintuitive relationships like those depicted in Figure 1 do not require that the sources of individual differences are entirely independent from those driving within-subject effects. They might still arise if there are too many potential interacting mechanisms contributing to individual variability, but only some of which have been exploited and understood through experimental manipulations (Borsboom, Kievit, Cervone & Hood, 2009). This problem of relating multiply determined individual differences to within-subject experimental manipulations has been most commonly discussed with regard to multifactorial traits such as personality and IQ (Cronbach, 1957; Borsboom, Kievit, Cervone & Hood, 2009). However as alluded to in Borsboom et al.'s quote ("barring perhaps the most basic laboratory tasks") the simpler the behaviour, the more likely it is to be strongly associated with one particular cognitive mechanism. In this case, it might be true that individual differences in task performance reflect the same underlying mechanism as studied in the experimental literature.

Historically, Hull (1945) expressed a related view when he argued that there are common laws governing behaviour across individuals and even species. If an experimentally derived behavioural 'law' or equation - for example, relating stimulus strength to behavioural outcome - has variables (such as stimulus strength) and 'constants', then Hull argued that while experimentalists manipulate the 'variables' it is in the 'constant' parameters that individuals and species will differ. Those parameters would represent 'primary' individual differences, which then combine and interact to produce the 'secondary' differences measured in overt behaviour. It would be almost impossible to derive these 'laws' (i.e. understand basic cognitive mechanisms and derive functional models) from only studying individual differences. However, if you already know the laws/mechanisms from experimentation, and have tasks that can specify the parameters for different individuals, then you could understand individual differences with respect to these mechanisms.

This, as we perceive it, is the basic hope and implicit assumption of the various ventures to use individual differences in well-known cognitive tasks to relate brain structure, clinical symptoms, or genetic variation to basic cognitive mechanisms. However, we are aware of no explicit investigation of whether this implicit assumption holds even for tasks that are considered to tap the most basic mechanisms. It is possible that individual variance still arises from a combination of sources, and that the main variability is independent from the mechanisms that are experimentally manipulated. Here, we focus on the relationship between the visibility of a prime stimulus and the effect it has on motor control.

**The case of visibility and motor priming**

It has become widely accepted that our 'voluntary' purposeful actions should rather be regarded as an interaction between processes occurring within and without the scope of conscious awareness (e.g., Aglioti, DeSouza, & Goodale, 1995; Boy, Husain, & Sumner, 2010; Neumann & Klotz, 1994). An important and perennial question arises from this framework: what is the relationship, if any, between our conscious awareness of a stimulus and the way in which that stimulus influences motor plans? A practical way to investigate non-conscious motor influences has been implemented in the masked priming paradigm (e.g. Leuthold & Kopp, 1998). Generally, prime stimuli speed responses to subsequent "target" stimuli if they are associated with the same response (compatible) and slow responses if prime and target are associated with different responses (incompatible). This positive compatibility effect (PCE) has been taken to demonstrate that a prime can partially activate the response associated with it, even though the participant had no intention of responding to the prime and may not even have perceived it (note that for priming to occur, the participant must be intending to respond to the target).

In some masked priming paradigms, a counterintuitive negative compatibility effect (NCE) has been measured, such that responses are faster and more accurate for incompatible primes than for compatible primes (for reviews, see Eimer & Schlaghecken, 2003; Sumner, 2007). Most interestingly for the present purpose, in the initial studies of the NCE, the direction of priming appeared to depend on whether the prime was above or below perceptual discrimination threshold: visible primes produced PCEs while invisible primes produced NCEs (Eimer & Schlaghecken, 2002; Klapp & Hinkley, 2002).

There are theoretical reasons to expect a strong relationship between prime visibility and priming effects. Firstly, the original theory of the NCE (Eimer & Schlaghecken, 1998, 2002, 2003), proposed that the very role of the inhibitory mechanism indexed by the NCE was to suppress weak motor activation evoked by previously learnt stimulus-response associations, unless the stimulus reached conscious awareness. Presumably stimuli that reach awareness are more likely to be behaviourally relevant than those that do not, and so it might be efficient to allow positive motor priming by stimuli we are conscious of, but to suppress it when we are not conscious of its source.

More generally, even if there were not any causal connection between visibility and the priming effect, we would expect a strong relationship if the same cascaded visual processing underlies both. We assume that some visual information gets through to motor areas regardless of whether it has reached conscious threshold – this is what causes subliminal motor priming – and we also assume that as the representation strength of visual information increases, then it is more likely to be consciously perceived. If we further suppose, not unreasonably, that increases in representation strength will also have a systematic effect on priming, then we predict there would generally be a strong correlation between priming and visibility even without any direct causal connection between them.

**Perception separate from action?**

Alternative to the above theoretical frameworks are proposals that pathways subserving non-conscious processes are distinct from those serving conscious awareness, or, relatedly, that pathways linking vision to action are distinct from those leading to perceptual experience (Milner & Goodale, 1995). It is also possible that differences in temporal dynamics, rather than simply anatomical pathway, constitute an important distinction between information driving priming and information supporting conscious perception. For example, priming may be mainly driven by an initial transient forward sweep of information, while awareness is supported by subsequent sustained recurrent processing (Bompas & Sumner, 2008; Lamme & Roelfsema, 2000). These 'dissociation' accounts do not require a correlation between visibility and priming.

As mentioned above, for the NCE paradigm, a strong relationship between visibility and the direction of priming was initially reported: invisible primes produced NCEs while visible primes produced PCEs (Eimer & Schlaghecken, 2002; Klapp & Hinkley, 2002). However, this relationship soon became controversial: NCEs were also found to occur when prime discrimination was above chance (e.g. Klapp, 2005; Klapp & Hinkley, 2002; Lleras & Enns, 2005; Mattler, 2006; Sumner, Tsai, Yu, & Nachev, 2006), and conversely, PCEs could occur with invisible primes (e.g. Lleras & Enns, 2006). It was also found that teaching subjects to discriminate the prime stimuli through extensive practice did not alter the NCE, which appears to rule out a causal relationship (Schlaghecken, Blagrove, & Maylor, 2008; we have replicated this in unpublished work).

However, none of these results necessarily mean there is no relationship between visibility and the NCE. Learning to perform better on prime discrimination presumably relies on learning about the subtle clues present around the time of the onset of the mask. One way to do this would be to learn what to attend to. Such attentional learning might not transfer to make primes more visible in the masked priming blocks, because here the participant is instructed to ignore the primes and must pay attention to the target. Moreover, as outlined above, there does not need to be a causal role for visibility in reversed priming for us to expect a behavioural relationship between the two - it would emerge if common visual processing leads to visibility on the one hand and priming on the other. With regard to findings that in some experiments visible primes can produce NCEs while in other experiments invisible primes can produce PCEs, this overturns a categorical distinction between invisible and visible primes, but it does not mean there is no relationship when other factors are held constant. Indeed, it is now generally accepted that there are at least two mechanisms that can contribute to NCEs, depending on the type of mask used (Boy, Clarke, & Sumner, 2008; Jaskowski, 2008; Klapp, 2005; Schlaghecken & Eimer, 2004; Sumner, 2008), and these may differ in their relationship to prime visibility. Overall, there are many indications in the literature that the more visible the prime, the more likely it is to produce a PCE, and conversely, NCEs are most easily produced with less visible or invisible primes (Klapp, 2005; Klapp & Hinkley, 2002; Lleras & Enns, 2004, 2005; Schlaghecken & Eimer, 2006; Sumner et al., 2006). However, like the counter-examples, these indications are not always free from other factors known to influence priming, such as differences in the mask.

**Experimental manipulation of prime visibility vs. individual differences**

Here we investigate two related questions: Firstly, is there a consistent relationship between average prime visibility, as measured by prime discrimination performance, and average priming effect when prime visibility is manipulated in different ways (prime contrast, prime-mask interval, mask contrast, mask density)? It is plausible that the relationship could depend on the way the stimuli are manipulated, because the different manipulations have different impacts on the initial visual burst of activity produced by the prime and the subsequent interaction of the prime activity and the mask activity. For example, if priming is primarily driven by the initial feed-forward sweep of prime activity, it may be more directly influenced by manipulating the contrast of the prime, than by manipulating the contrast of the mask, even though both affect prime visibility.

Secondly, across participants, is there a systematic relationship between an individual's priming effect and how well they can discriminate the prime? We would expect this to be the case if there is a causal relationship between visibility and priming, or if priming and visibility depend on the same perceptual representations, as discussed above. More generally, with the resurgence of interest in individual differences in perception and cognition, it seems to be a common assumption that established within-subject phenomena will have easily interpreted translation to individual differences. But as discussed at the beginning of the Introduction, a correlation between priming and visibility across people does not necessarily follow from a relationship between average priming and average visibility produced by manipulating the stimuli (question 1).

Several previous studies have assumed a correlation in their methodology, because they have used a individual's prime discrimination ability to set the 'appropriate' prime strength for that participant in priming blocks (e.g. Boy & Sumner, 2010). This approach appeared to 'work', but was never directly compared to using the same prime strength for all participants. In other experiments, both the presence (Eimer & Schlaghecken, 2002; Klapp & Hinkley, 2002) and absence of a correlation has been reported (Hermens, Sumner, & Walker, 2010), but whenever the correlation has been measured with a single set of stimuli – i.e. with one particular prime strength – it could be misleading due to the biphasic nature of the priming effect. We will explain this in due course, along with the approach taken to overcome this problem, based on Eimer and Schlaghecken (2002).

## 2. Methods

**Overview of experiments**

We present five new experiments and reanalyses of two previous ones, making seven sets of data with which to answer the two questions set out above. Experiments 1 and 2 use prime duration and prime brightness, respectively, to manipulate visibility. The previous data sets, from Sumner et al. (2006), used the same manipulations but with fewer degrees of visibility, and thus represent replications of Experiments 1 and 2. For the logic of exposition they are therefore branded as Experiments 3 and 4. The original purpose of these previous experiments was to investigate the effect of attention on priming, but since the attention effect is not at stake here, we average over the attentional manipulation. Experiments 5 and 6 manipulate the visibility of the prime by playing on the properties of the mask, modulating its brightness or its density, respectively. Experiment 7 manipulates prime brightness like Experiments 2 and 4, but implements a different way of determining prime visibility exactly following Eimer and Schlaghecken (2002).

a. Participants (all experiments)

62 participants (48 women; age 18–38) from Cardiff University participated in five experiments (respectively 11, 10, 10, 11 & 20 participants). Details about participants for Experiments 3 an 4 are to be found in Sumner et al. (2006). All self-reported having normal or corrected-to-normal vision, no history of brain damage and were right-handed.

b. Apparatus (all experiments)

Stimulus presentation was performed by a PC-controlled Cambridge Research Systems (CRS) Visage® connected to a 21" Sony GDM-F520 Trinitron monitor. Stimulus presentation was synchronized with the screen refresh rate of 100 Hz, and timings were controlled and measured by the CRS clock and thus not subject to the errors produced by normal PC operating systems. Manual responses were collected using a CRS-CB6 button box.

c. Masked-priming task

The protocol is given in full for Experiment 1, and deviations from this in the other experiments are detailed below. Participants made speeded responses with a left- or right-hand key press (counterbalanced) to right and left arrows (1° x 1.5°), which occurred in random order and located at 0.5° from fixation, in a random direction from fixation (see Fig. 2-B). A fixation cross was visible at the center of the screen at the beginning of each trial. The primes were identical to either one or the other target, but presented for various duration between 10 and 60 ms (by steps of 10 ms), and appeared within 0.5° of fixation (i.e., in the same vicinity as the target, but not in an identical location on any trial). In all conditions the prime was followed by a 100 ms mask of 2.2° x 2.2° and constructed of 36 randomly orientated lines, excluding any orientation closer than ± 5° to the orientations present in the prime and target stimuli. A new mask was constructed on each trial but appeared always in the same place, centred on fixation. All trials had a long mask-target SOA of 150 ms. The background was dark grey (10 cdm$^{-2}$) and other stimuli were light grey: fixation cross, primes, targets and masks were 60 cd/m$^2$. 480 trials were presented in a random order (60 trials for each prime duration), with brief breaks every 60 trials. Participants were not informed about the different prime durations. Stimulus sequence illustrated in Fig 2-B.

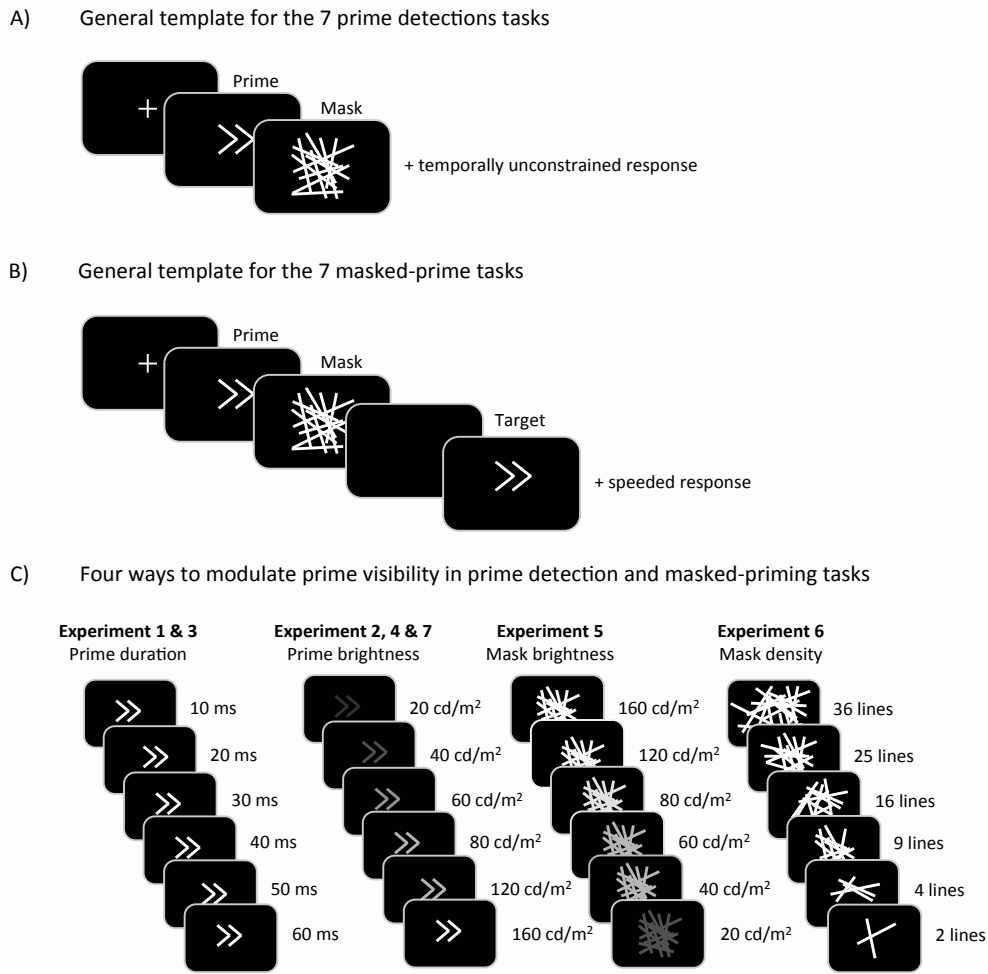

Figure 2: A) Illustration of the stimulus sequence in the prime detection tasks. (B) Illustration of the stimulus sequence in the masked-priming tasks. C) Illustration of the four ways to alter prime visibility, with primes getting stronger or masks getting weaker from top to bottom. Because prime visibility is changed through the modulation of the physical characteristics of the prime or the mask, we will refer to changes in the prime 'strength'.

d. Prime identification tasks

In experiments 1 to 6, prime visibility thresholds were assessed individually before or after each masked-priming task using a procedure of constant stimuli (240 trials in total, testing each of the prime or mask conditions in randomized order). The participants' task was to guess the identity of the prime (forced choice). For each experiment, stimulus sequence and timing was identical to the masked priming protocol, but with the target omitted (Fig 2-A). In Experiment 7 instead of using a method of constant stimuli to assess prime visibility, we used a 2-up 1-down staircase procedure following Eimer and Schlaghecken (2002). The staircase continued until four consecutive staircase reversals and data from the staircase were then used to plot the individual's prime visibility psychometric function. Discussion of the procedure (particularly why our procedure does not contain the target stimulus) can be found in a previous publication (Boy & Sumner, 2010).

e. Experiment 2

The procedure was identical to Exp. 1 except that we fixed the duration of the prime at 40 ms and manipulated its brightness to cross the perceptual threshold. Six levels of brightness were selected (20, 40, 60, 80, 120 & 160 $cd/m^2$) and picked in a randomly shuffled order on each trial (see Fig 2-B).

f. Experiments 3 & 4

The datasets termed experiment 3 and 4 are re-analyses of the data collected by Sumner et al. (2006). As their dataset also contained an attentional cue manipulation that is not of interest for the present purpose, we collapsed the validly and invalidly cued conditions. Experiment 3 reanalyzes the data in their second experiment for primes lasting 20, 30, 40, 50 and 60 ms. Experiment 4 reanalyzes their third experiment in which they used 4 levels of prime brightness (20, 40, 80 & 160 $cd/m^2$). Note that in both these experiments, presentation was blocked (256 trials for each prime duration or each prime brightness), rather than randomly shuffled, as in Experiments 1 and 2 above.

g. Experiments 5 & 6

In these experiments, prime duration was fixed at 40 ms and the manipulation affected the brightness of the mask (values between 20 & 160 $cd/m^2$) or its density (mask composed of 2, 4, 9, 16, 25, 36 lines, at random in any of the grid positions, see Figure 2-B). On each trial, one of the six levels of mask brightness (Expt. 5) or mask density (Expt. 6) was selected in a randomly shuffled order (a total of 80 trials for each of the 6 levels of brightness or density). Other details are as in Expt 1.

h. Experiment 7

The masked-priming task in Experiment 7 was identical to Experiment 2 where prime brightness was manipulated (in 6 levels, 20, 60, 60, 80, 120 & 160 $cd/m^2$). The difference resides in the prime identification task, which utilized a staircase procedure (see above). We also ran more participants.

## 3. Results

a. Average compatibility effects

For average compatibility effect (CE, average reaction time to incompatible trials - average reaction time to compatible trials), the results are unambiguous and simply stated. We find a clear transition from negative CEs for weak primes to positive CEs for stronger primes in all experiments (all Fs > 15, all *p*s < .0001, see Fig 3). The pattern is independent of the way prime strength was manipulated. Since visibility (prime discrimination) also rises systematically with these manipulations of the stimuli, strong correlations occur in each experiment between CEs and prime discrimination for each stimulus condition for each participant (all rs > .56, *p*s < .001; see Fig 4). However, as we shall see below, these correlations are entirely driven by the manipulation of stimulus strength, and not at all by individual differences in discrimination ability.

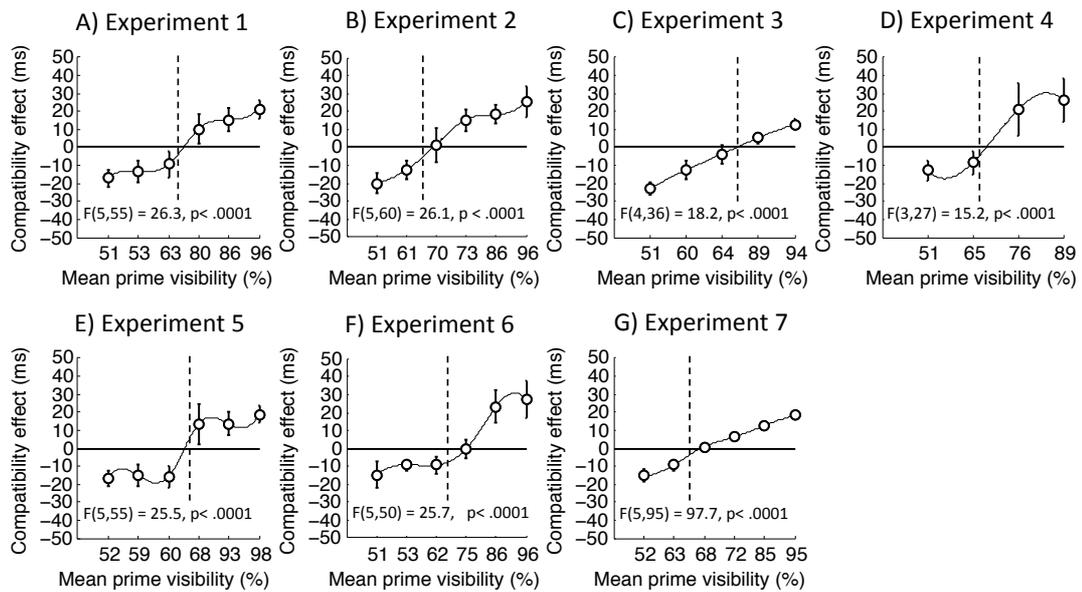

Figure 3. (A:G): Average compatibility effect as a function of mean prime visibility for each level of prime visibility for all seven experiments (error bars represent the inter-individual standard error of the mean). A spline fit connects the data points (we have no theoretical basis for any particular curve fit). As in Eimer and Schlaghecken (2002), the dotted vertical line on each plot indicates the 66% prime discrimination threshold along the prime visibility gradient as determined in the prime identification tasks.

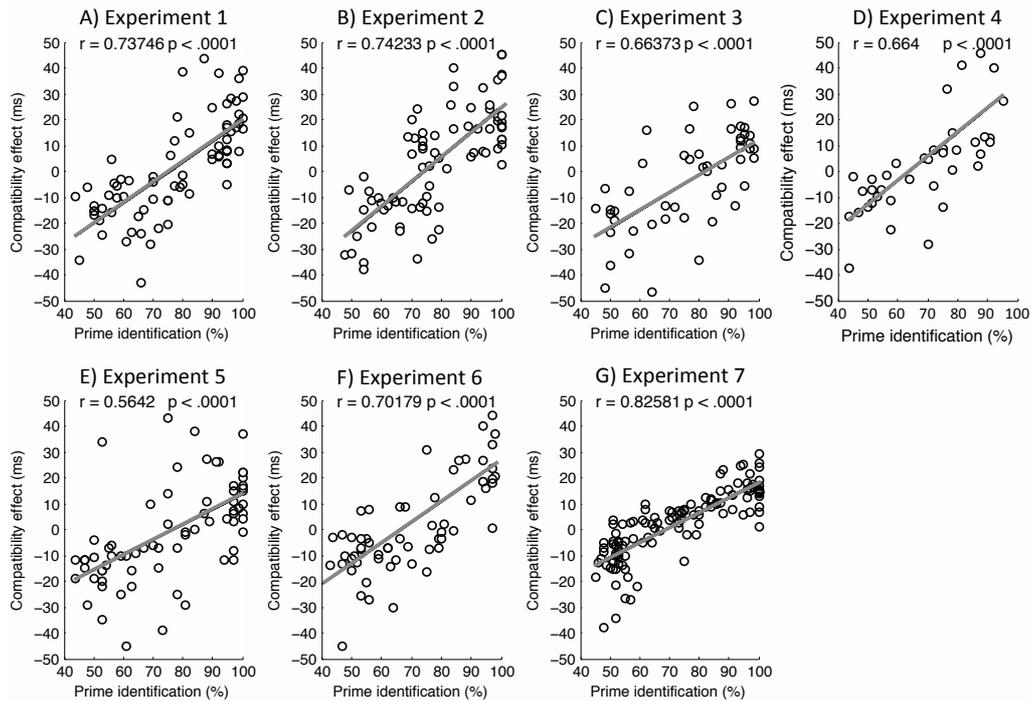

Figure 4. (A:G): Scatter plots of the compatibility effect against prime discrimination. The coefficient of correlation of the linear regression is presented (along with its p-value).

b. Correlations between priming and visibility across individuals

*Approach:* If priming is zero with no prime, negative with an intermediate strength prime and positive with a strong prime, there are two places in this relationship where it is weakly negative: with very weak primes, or with stronger primes that are not quite strong enough to produce positive priming. Measuring the simple correlation between discrimination and priming could therefore be misleading, because subjects showing the same level of priming could actually be at different points on the biphasic relationship between prime strength and priming effect. To get around this problem, we followed Eimer and Schlaghecken (2002) and took the approach of measuring both discrimination and priming effect for multiple prime strengths. Then, for each participant, the discrimination threshold was extracted from the psychometric function of discrimination performance against prime strength, and the priming transition point (negative to positive) was extracted from the curve of CE against prime strength (see Fig 5). In other words, for each individual, we found the prime strength for which discrimination accuracy was 75%, and the prime strength for which negative priming turned to positive priming (the zero crossing). If prime visibility is related to priming, these two measures are expected to positively correlate.

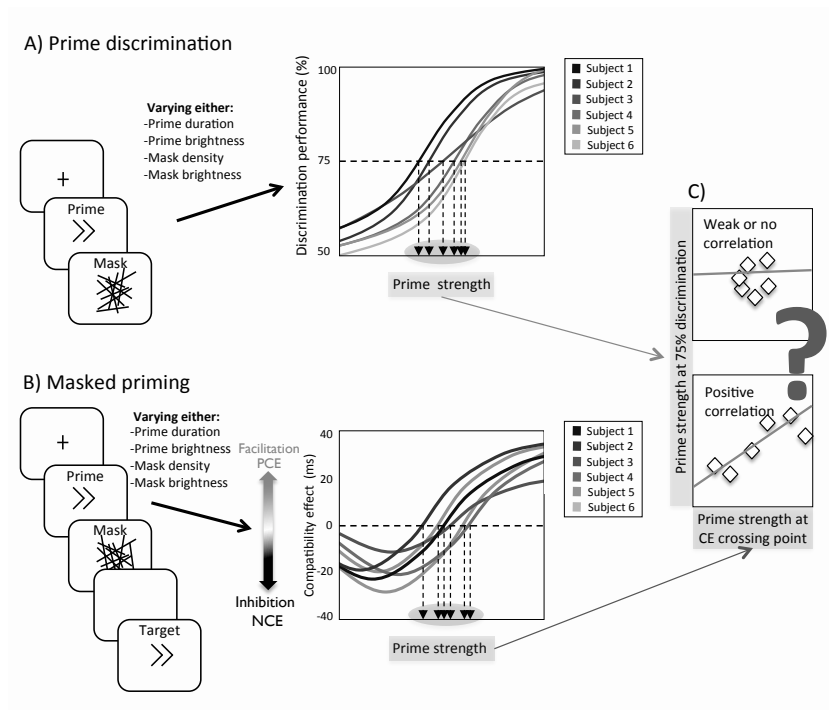

Figure 5. Illustration of the data processing chain used for testing correlation between individuals' priming effect and how well they see the primes. A) Prime discrimination: derivation of the prime strength at which each subject shows 75% discrimination of the prime stimuli (which corresponds to guessing in 50% of trials). B) Masked priming: derivation of the prime strength for which each subject's compatibility effect goes from negative to positive. C) Correlation between these two measures.

In seven experiments we found only one hint of a positive correlation (Experiment 1, r = .28, p= .38). In all six other experiments, correlations coefficients ranged from -.21 to -.07 (all ps = NS, Fig 6). Of course, null results are difficult to be sure of. One possibility, especially with relatively small N, is that one or two outlying values can destroy the statistical correlation even though a true correlation might exist. To assess the likelihood of this, we used jackknife estimates, which, reassuringly showed that none of the correlation coefficients are likely to be affected by a large bias; on average, the "true" coefficient of correlation was probably misestimated by not more than ± 0.09 (Table 1, row 2).

Another possibility is that it is possible that any given subsample of a population, by chance, will not show a correlation even though a correlation exists in the whole population. We used simulation to estimate the chance of this happening in all seven experiments for the N we used in each experiment. Only one experiment in a total of seven found a numerically positive r-value (but statistically not significant). Thus we estimated the chance that despite there existing a real between-subject correlation between visibility and priming, we only obtained once a positive r-value in all seven experiments. To do this we assume that despite any noise in our variable measures, if we had measured a sufficiently large number of participants we would have revealed a correlation if it exists. For exposition, let us assume it would be r=0.3. We therefore simulate a large population (100 000 data points) with a correlation of r=0.3. We then simulate each experiment by randomly selecting the same number of points as we had participants in that experiment, and we calculate the r-value we obtain with this subsample. We repeat this for each of the seven experiments, and then count the number of positive and negative r values obtained. We repeat this procedure 100 000 times, to obtain the probability that only one (or less) of our seven experiments would give an r value above zero, if the real r value is, for example, 0.3.

We repeated this for 'real' r-values from 0 to 1. For 'real' r-values of 0.4 and above, the simulation produced zero occurrences of our data pattern in 100 000 iterations. For a real value of r=0.3 there were 16/100000, for a real value of r=0.2 there were 174/100000, for r=0.1 there were 1352/100000, and for r=0 there were 5914/100000. In other words, the probability of getting our results for a real r>0.3 is very low, and the probability for a real r=0.2 is about 30 times (5914/174) lower than if the real r-value is zero (for real r=0.1 it is about 4-5 times lower than for no correlation). From this we conclude that the likelihood of there being any sizable correlation (r>0.2) between visibility and priming is very small, given our data.

We also checked whether the results were specific to extracting discrimination threshold using 75% performance. They were not; using either a 66% or a 70% threshold also produced no hint of positive correlation for experiment 2 to 7 (-0.18< $r$ < 0.09). r-value for experiment 1 stayed close to the estimate at 75% performance (respectively .24 and .27). This is important because participants can differ in the slope as well as the position of their psychometric functions, and thus their rank order can change if we use different criteria for what their conscious threshold is. Note that at 75% discrimination accuracy, participants are 'seeing' (or basing their answers on information) 50% of the time, since that pure guessing would produce 25% correct answers when there are 25% incorrect answers. At 66% accuracy, they are 'seeing' the stimulus 32% of the time (100-34*2). Thus the range of visibility levels we have used for the analysis spans 32%-50% seen targets, which we believe appropriately reflects the transition from subliminal to supraliminal. Note that the NCE to PCE transitions are also in this region (Figure 3). In case the correlation is better reflected by visibility performance near chance levels for the very weakest primes, we also correlated the earliest points on the psychometric functions (where they cross the y-axis) with the CE transition points. Of the seven experiments, we only obtained hints of positive correlation for experiments 1 and 7 (respectively .34 and .23, ps= NS). Finally we performed a further analysis that does not rely on estimating correlations at all (see section c).

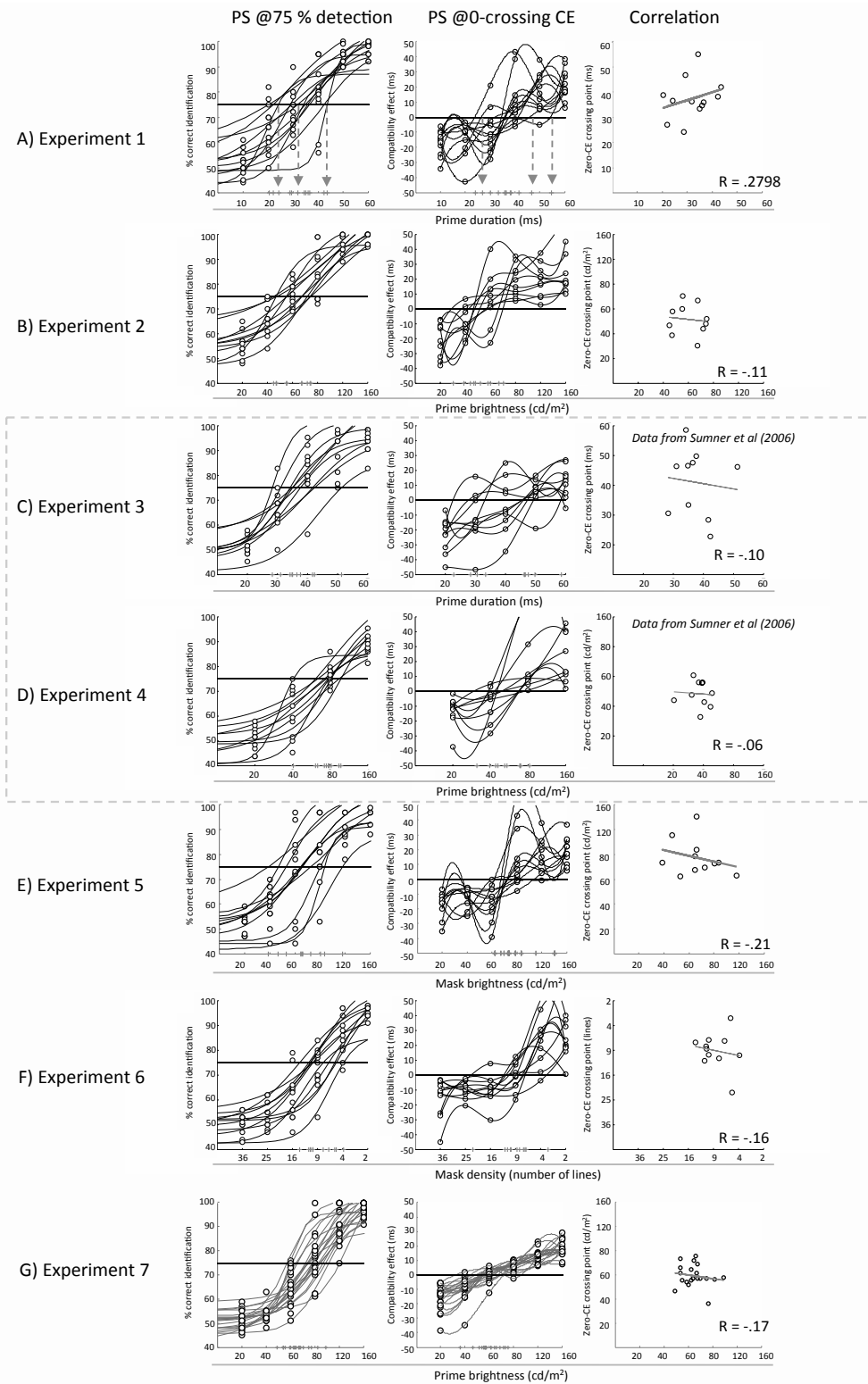

Figure 6. (A:G): Individual data for the six experiments. Column 1: Prime strength at 75 % detection thresholds in the prime detection task. Column 2: Prime strength at the Zero-CE. Column 3: Scatter plot of detection threshold against priming transition point. The red arrows in row A show examples of how these values are derived (also refer to Figure 5).

|  | Expt. 1 | Expt. 2 | Expt. 3 | Expt. 4 | Expt. 5 | Expt. 6 | Expt. 7 |
|---|---|---|---|---|---|---|---|
| Pearson's r | .245 | -.1061 | -.105 | -.0616 | -.2167 | -.1605 | -.1671 |
| Jackknife bias est. | -0.002 | -0.003 | -0.06 | -0.09 | -0.01 | 0.06 | 0.015 |
| Z-score difference to Eimer & Schlaghecken | 0.26 | 2.07 | 2.06 | 1.97 | 2.43 | 2.30 | 2.35 |
| p-value | .441 | .047 | .048 | .057 | .021 | .029 | 0.024 |
| Boot-strap p-value | .104 | .037 | .031 | .042 | .0056 | .029 | 0.021 |

Table 1. Rows 1 & 2: Pearson's coefficients of correlation and their Jackknife bias estimates for the seven experiments. Rows 3 & 4: Results of the Fisher's Z-score difference test comparing correlations obtained in the seven experiments to that calculated for Eimer and Schlaghecken's (2002) datasets. Row 5: P-values derived from bootstrapping approach. See sections b and d of results for details.

c. Better and worse discriminators.

As a further analysis to test whether there is any effect of discrimination ability on priming, we plotted average CE curves for two groups of participants in each experiment – those with above median discrimination scores, and those with below median scores (we did this based on the individual 66% threshold values). Although this approach is more blunt than approaches taken above, it has two advantages: it includes all the participants, whereas above we could not include participants if their CE curve did not cross from negative to positive; it presents a clear visualization of whether the priming curve depends on prime visibility without the need for a further, less intuitive, analysis step (see Figure 7). We found no hint of any effect of visibility on the CE curves.

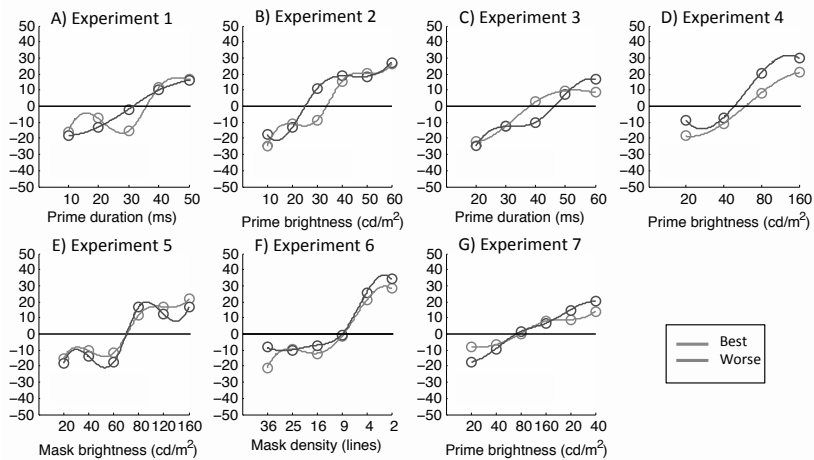

Figure 7. (A:G): Average CE curves for participant demonstrating best and worse discrimination in each experiment – those with above median discrimination scores, (red – labeled "Best") and those with below median scores (blue – labeled "Worse") ) A spline fit connects the data points (we have no theoretical basis for any particular curve fit). If the priming effect became positive only as an individual was better able to discriminate the prime, we would expect the blue curve to be shifted to the right relative to the red curve, because worse discriminators would need more powerful primes (relative to the mask) to produce positive priming. This is clearly not the case.

d. Direct comparison to Eimer & Schlaghecken (2002)

Eimer & Schlaghecken (2002) used the same framework as we have done, but unlike us, they found positive correlations between discrimination threshold and CE crossing point in two experiments, which manipulated prime strength using prime duration and mask density. To make a direct comparison with our data, we pooled together data from the two experiments by Eimer & Schlaghecken (2002) by Z-scoring them and obtained a correlation of r = .67 (p< .0007) for this new combined set of data. We also transformed the r-values from our experiments into Z-scores (through the r-to-z transformation method defined in Fisher, 1915): In five out of seven experiments, our correlation differed significantly from that of Eimer and Schlaghecken (see Table 1, 3$^{rd}$ & 4$^{th}$ row). Further, because there is possible bias in this parametric comparison introduced by distribution distortion in small samples, we used a data-driven bootstrapping approach. We derived a distribution or r-values for each experiment (and for the combined data in Eimer and Schlaghecken, 2002) by resampling the data (with replacement) 10000 times. The overlapping area under any two distributions then gives the p-value for the null hypothesis that the two r-values do not differ. These are given in Table 1, 5$^{th}$ row, for the comparisons of each of our experiments with the combined data of Eimer and Schlaghecken's two experiments. In Experiments 1-6 we used a method of constant stimuli to provide the psychometric function for estimating prime visibility. Eimer & Schlaghecken (2002) used a staircase procedure. In case this might make a difference to participant's behaviour, Experiment 7 used a staircase procedure. The results were indistinguishable from Experiments 2-6, and significantly different to Eimer & Schlaghecken (2002) (see Table 1) suggesting that the exact method of assessing visibility is not responsible for our failure to find a correlation.

## 4. Discussion

Our results lead both to specific conclusions about the disputed relationship between visibility and reversed masked priming, and also more general conclusions about how relationships that are clearly apparent across stimulus manipulations can be entirely absent across individual differences. The latter issue is of general interest for the growing use of cognitive and sensorimotor tasks to study individual differences.

**Implications for the study of individual differences.**

Despite the theoretical problems, outlined in the Introduction, of relating individual differences to experimental manipulations (e.g. Borsboom, 2006), in practice there are many examples of cognitive tasks that were developed within the sphere of within-subject designs being widely employed in the study of individual differences. This trend is increasing with the rise of genetics and the search for cognitive endophenotypes of psychiatric disorders, for early markers of dementia/cognitive impairment, and the development of ever more sophisticated brain imaging techniques that are analyzed at the individual level, rather than using group-averages (voxel-based morphometry, diffusion tensor imaging, magnetic resonance spectroscopy, dynamic causal modeling etc.) With respect to these growing fields, our results can be taken as an exemplar cautionary tale: even relationships between simple tasks and basic cognitive constructs that are well worked out in the realm of controlled stimulus manipulations and within-subject designs, may not transfer to individual differences. Just as has been statistically pointed out (e.g. Borsboom, 2006, Borsboom et al, 2009), and as illustrated in Figure 1, the inter-individual variance can have an entirely different source from that produced by stimulus manipulations.

The problem is likely to arise due to the multiple factors contributing to individual differences, even in simple tasks. For example, various 'inhibition' tasks have been employed to study self-control and impulsivity, such as the antisaccade, Stroop task or stop-signal task. For these tasks, it is generally assumed that individual differences in performance reflect basic differences in inhibition ability, but this simple conclusion is actually not well supported by the fact that performance tends to correlate only very poorly across tasks supposed to measure the same inhibition ability (Barch, Braver, Carter, Poldrack, & Robbins, 2009; Cothran & Larsen, 2008; Cyders & Coskunpinar, 2011; Friedman & Miyake, 2004; Schachar, 2011). Thus just like in complex traits such as IQ and personality, individual differences in relatively simple behavioural tasks may also be too multiply determined to be easily matched to the cognitive mechanisms revealed by experimental manipulation.

This is not to say that there is no point attempting to relate individual differences in cognition to the mechanisms explored through experimental manipulations. Rather, we argue that the endeavor will be more fruitful if approached with the understanding that the relationship will be complex and tricky to work out, rather than the assumption that individual differences automatically reflect the same mechanisms studied by within-subject manipulations in the same task. We recognise that this implicit assumption is very hard to avoid, and we have made it ourselves previously. Even Borsboom et al (2009) in their excellent exposition of the statistical problem of relating inter- and intra-individual variance, appear to conflate between and within participant effects at one point (p. 24-26, a study of differences in chess playing between expert and non-expert groups is used to support a conclusion about gaining expertise within individuals).

That even the best statisticians succumb occasionally appears to confirm the deeply intuitive and appealing nature of such conclusions - which may sometimes be correct. But given their intuitiveness, the upmost vigilance will be required to work out when such assumptions are unfounded. This can be supported by the adoption, where appropriate, of statistical techniques that allow within and between participant variance to be analysed simultaneously (e.g. multilevel modeling, nested design, see Snijders & Bosker, 2012; Kliegl et al, 2011).

Furthermore, different sources of inter- and intra- individual variance can be exploited to form more stringent tests of theory. For example, if X is hypothesized to cause Y, then there should be a systematic relationship between X and Y both in intra- and inter- individual variance. If both variances are assumed to come from the same source this might be seen merely as a replication, which would not encourage researchers to assess both.

**The relationship between prime visibility and the NCE.**

The role of visibility for the 'inhibitory' component of masked priming – the NCE – has been disputed for a decade (for reviews, see Eimer & Schlaghecken, 2003; Sumner, 2007). In essence, there have been *two* questions at stake: Firstly, whether there is any systematic relationship between prime strength / visibility and the direction of priming; secondly, whether there is a causal connection between awareness and the occurrence or not of motor inhibition. Initial studies found that primes presented below the threshold of conscious visibility were categorically associated with NCEs whereas visible primes were associated with PCEs (Eimer & Schlaghecken, 2002; Klapp & Hinkley, 2002). Other studies implied that the transition from NCE to PCE seemed to occur in a continuous manner: as the prime got more visible, priming got more positive (Klapp, 2005; Schlaghecken & Eimer, 2006; Sumner et al., 2006). However some studies found – and some authors strongly argued for – no association between prime visibility and the direction of priming (Jaskowski, Bialunska, & Verleger, 2007; Lleras & Enns, 2004; Verleger, Jaskowski, Aydemir, van der Lubbe, & Groen, 2004). There have often been difficulties of interpretation because prime visibility is normally confounded with changes to the stimuli – such as the masks – which are thought to have their own effects on priming (Jaskowski, 2008; Jaskowski et al., 2007; Lleras & Enns, 2004; Verleger et al., 2004) and conversely, studies aiming to investigate different masks have often been confounded by visibility differences. Schlaghecken et al. (2008) circumvented this problem by changing prime discrimination through perceptual learning without changing the stimulus. Improved prime discrimination ('visibility') did not correspond to any change in the priming effect, providing the strongest evidence yet against a causal role for visibility. However, it remained possible that discrimination had improved through participants learning to attend better to the small cues available to guide prime discrimination. Such learning might not have transferred to masked priming blocks because participants now had to ignore the primes and attend to the target.

Our approach was to test whether any relationship consistently held across different types of stimulus manipulation, and across individual ability. We found a clear systematic relationship between visibility (prime strength/mask weakness) and the direction of priming in every dataset (Fig 3, 4, 7). Since this held for four different ways to manipulate visibility (prime duration, prime brightness, mask brightness and mask density), it is unlikely simply to reflect one type of stimulus characteristic. It appears to be a more general product of the relative strengths of prime and mask. Thus in answer to the first question - is there a systematic relationship? - we conclude that under most types of prime or mask manipulations, there is a strong relationship between prime visibility and priming.

To further test the causal hypothesis, we took advantage of an alternative source of variance in visibility – individual differences. As in the perceptual learning approach of Schlaghecken et al. (2008), this is also not confounded by stimulus changes. Here we found, across all seven datasets, no hint of any positive correlation between priming and an individual's ability to discriminate the primes (Fig 6 & 7), supporting the conclusions of Schlaghecken et al. (2008) that there is no causal influence. We used converging approaches to ascertain whether this was simply due to lack of power or outliers in the data (jackknifing, simulation, bootstrapping, direct comparisons to the previous study of Eimer and Schlaghecken, 2002, and grouping the CE data by a median split of the discrimination data, Fig 7). We cannot explain why our results do differ from those of Eimer and Schlaghecken (though we speculate below), but we can appeal to weight of evidence (seven datasets here, plus the evidence from Schlaghecken et al., 2008, vs. two experiments in Eimer and Schlaghecken, 2002). Further, it is essential to note that the logic that our framework for understanding the data, spelled out in the following sections, does not disallow correlations to occur – indeed we have shown they clearly occur for within-subject stimulus manipulation – they just do not reflect direct causal linkage when they do.

**Implications for theories of the NCE**

Our results are inconsistent with the original theory of the NCE (Eimer & Schlaghecken, 1998; Klapp & Hinkley, 2002), because it contained a causal role for prime visibility, proposing that automatic motor inhibition occurred as a result of the initial prime-related motor activation failing to reach awareness. The results are consistent with two later theories, which both emphasize the importance of the mask stimulus, but do not envisage a causal connection between prime visibility and the NCE.

The 'object updating' (or 'active mask', or 'mask-induced priming', Lleras & Enns, 2004, 2005; Verleger et al., 2004) account suggested that the NCE was caused not by motor inhibition, but by positive priming in an unexpected direction due to prime-mask interaction. When the mask contains elements of both possible primes, then the prime-mask sequence can also be considered as a sequence of both primes presented overlaid with a brief temporal separation. The 'prime' that appears second (i.e. the new elements of the mask) could then reverse any priming associated with the prime that appeared first (i.e. the actual 'prime'). In this case, either increasing the perceptual strength of the first prime, or decreasing the perceptual strength of the second prime (our mask manipulations) would be expected to increasingly favour positive priming over reversed priming, creating the systematic relationship we found. However, previous studies have shown that the object updating account is very unlikely to explain the NCE with the type of mask stimuli we employ, which are not made up of overlapping prime stimuli (Sumner, 2007). Therefore, we turn to the second mask-related theory.

The mask-triggered inhibition account (Boy et al., 2008; Jaskowski et al., 2007) shared the element of automatic motor inhibition with the original theory of Eimer and Schlaghecken, but it also focused on the mask like the object updating / active mask theory (Lleras & Enns, 2004, 2005; Verleger et al., 2004). It proposed that the inhibition must be triggered by a second stimulus that occurs after the prime – in this case the mask. New behaviorally relevant stimulus onsets are proposed to elicit inhibition of motor activity associated with previous stimulus – an automatic version of the bridge telling the engine room 'hold that last command, new information received…' In this case, it is clear that weaker masks might progressively weaken the triggered inhibition, and thus make an NCE less likely.

Why stronger primes should also make the NCE less likely requires a bit more discussion, since the NCE has been found to strongly mirror the positive priming effect measured at shorter prime-target intervals (Boy & Sumner, 2010). We might therefore expect that stronger primes would lead to stronger positive priming and also stronger inhibition. That this is not the case implies that the inhibition mechanism may be limited in strength, and once the initial positive deflection of motor activation becomes too strong, it cannot be fully reversed. This explanation is consistent with the arguments of Lingnau and Vorberg (2005), who pointed out that the occurrence of a PCE does not mean inhibition is absent – just that inhibition was insufficient to reverse the initial activity. After all, it is plausible that the functional role of such inhibition would not be to *reverse* the direction of motor balance, but to return it towards baseline. It may be that only within a tight range of parameters in an artificial laboratory situation do we find that the elicited inhibition over-compensates for the initial activation, creating the NCE. Note that for the sake of clarity, we have chosen to speak simply in terms of the relative *strength* of inhibition and activation mechanisms; we could also envisage that manipulations of the prime and mask differentially affect their response profiles across time.

**Implications for dissociation of visibility and sensori-motor mechanisms**

The lack of correlation across subjects, accompanied by the clear relationship across stimulus manipulation, allows us to go further than simply selecting a theory of the NCE that does not require causal connection between visibility and the NCE. Even the mask-triggered inhibition account (and the object updating account) would, at first sight, predict that if the relationship is present across stimulus manipulation, we would expect it across participants too. A participant who shows greater prime discrimination ability presumably has, in some way, a stronger representation of the prime relative to the mask than a participant who cannot discriminate the prime with the same stimulus settings. In other words, we assume that the direction of priming is caused by the relative strengths of activation and inhibition processes, which are in turn related to the relative strengths of prime and mask signals in the visual system. If we further assume that prime discrimination performance also reflects the relative strength of prime and mask signals in that person's visual system, then we should find the correlation between visibility and priming across participants as well as across stimulus manipulations (Fig 8-A). That we do not, tells us that inter-subject variance in discrimination must arise mainly from a difference source than inter-stimulus variance in discrimination.

The simplest solution to this would be that visibility and priming rely on entirely separate processes in different brain regions, that share only some initial early visual stage (Fig 8-B). Such conception of separate routes to process different aspects of a stimulus are not uncommon in psychology and echoes the famous dissociation between the processing of visual information for perception or for action (Milner & Goodale, 1995). If stimulus manipulations affect the shared visual stages, that would cause the relationship we reliably found. If inter-subject variabilities in visibility and priming arise mainly not in the early visual mechanisms, but in the further processes supporting awareness or motor processes separately, then there would be no correlation just as we found.

We speculate that the main locus of individual variation is not fixed, and will depend on study parameters and the idiosyncrasies of participants. If in some studies a substantial portion of inter-individual variance happens to arise from the shared visual stages, a correlation will be found between visibility and priming, just as reported by Eimer and Schlaghecken (2002).

Anatomical separation is not required, however, to explain our results, and neither is a view that some visual processes are 'for' perception while others are 'for' action. We believe that all vision and perception is, in some sense, for action, but there are different degrees of temporal immediacy between visual and motor mechanisms (Bompas and Sumner, 2008). For example, the temporal distinction between a feed-forward sweep and a subsequent phase of recurrent processing (e.g., Lamme & Roelfsema, 2000) could also provide our dissociation between stimulus manipulations and cross-participant correlation. Rapid motor priming and inhibition are presumably triggered by the feedforward sweep, while conscious perception is thought to rely on recurrent processing (Fig 8-B). Indeed, this is the main explanation for how subliminal priming is possible at all. Individual differences in recurrent processing need not correlate with individual differences in the feedforward sensorimotor sweep, leading to no correlation between visibility and priming. However, if stimuli are manipulated, then both feedforward and recurrent phases are necessarily affected, and hence a systematic relationship between visibility and priming occurs.

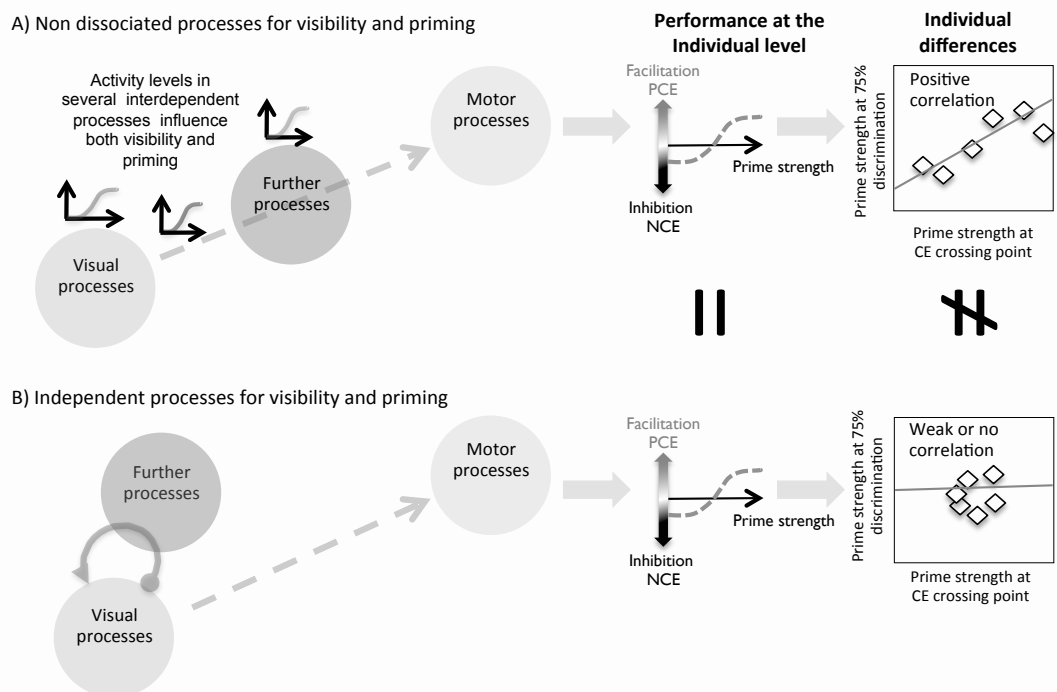

Figure 8: A) Even without direct causal influence of visibility on priming, if the ability to identify the prime and the level of positive or reversed priming arise from the strength of representations in the same cascaded processing pathway, a correlation is expected.

However, B) if the processes are separate, either anatomically or temporally, no correlation between them is expected.

**A final twist in the tale/tail of the relationship**

There is a final complication in the relationship between prime strength and the NCE, which recent evidence suggests is in fact consistent with the conclusions of this paper. When masked primes have been presented in the periphery, or when primes at fixation have been degraded, a PCE, not an NCE has occurred (Schlaghecken & Eimer, 2000, 2002, 2006). In other words, not only do strong primes produce PCEs, so do very weak primes (at least under some circumstances), and NCEs occur only for a band in between. To explain this, Schlaghecken & Eimer (2002) invoked a threshold mechanism by which inhibition is not triggered unless the initial prime-related activation is sufficiently strong (though still sub-motor threshold). Lingnau & Vorberg (2005) put this issue to test and systematically varied prime eccentricity, prime size and mask-target SOA, and argued that rather than a threshold below which inhibition does not occur, primes that leave weaker, smaller cortical representation could simply produce weaker inhibition with slower time course.

Interestingly, none of our seven datasets showed any hint of this PCE for the weakest primes. It is possible that none of our primes were weak enough, relative to the masks and targets we used. Additionally, it is likely that there is not a simple weakness metric for primes that differ on various dimensions. Although all our manipulations here behaved in effectively the same way with respect to the peri-threshold NCE-to-PCE transition, it remains possible that the PCE seen previously for very weak primes does not occur for all ways of making a prime weak. Previously, the PCE for weak central primes with around 150 ms mask-target SOA (as we had here) was produced by adding noise to very brief primes (Schlaghecken and Eimer, 2002). None of our manipulations emulated this procedure, and most of our primes were 40 ms long. When primes were shorter than this, they were high contrast and none were presented in the context of noise. Just as we find here that prime visibility does not straightforwardly predict the transition from NCE to PCE for stronger primes, a simple 'perceptual weakness' metric may not predict whether weak primes produce a PCE or not.

Consistent with this, Budnik, Bompas & Sumner (2013) recently reported that even when equated for visibility, peripheral and central primes still produced opposite priming effects. This indicates that there is no simple metric of perceptual strength between fovea and periphery that both predicts priming and is reflected by discrimination performance. Rather, there seems to be a distinction between a prime's ability to reach conscious awareness (which Budnik et al. (2013) called 'perceptual strength') and its ability to elicit motor activation and inhibition (which Budnik et al. (2013) called 'sensorimotor strength'). Such a distinction is fully consistent with Figure 8-B, and our finding that individuals' ability to see primes does not predict their priming effects.

**Conclusions**

We have found a reliable systematic relationship between prime visibility and the direction of priming when stimulus properties of prime or mask are manipulated, but we have also shown that this was accompanied by a consistent lack of correlation across participants. We conclude that the relationship across stimulus manipulation occurs due to the relative impacts of prime and mask signals on motor activation and inhibition, consistent with the mask-triggered inhibition account of the NCE. Individual variance in discrimination ability must arise from a different source, probably the recurrent processes that support awareness. In a more general context, the clear coexistence of correlation across stimuli, but not across people – and thus the fact that the respective variances must arise from different sources – has cautionary implications for the interpretation of even relatively simple cognitive tasks in the study of individual differences.